# Modelling the Dissipation Range of von Kármán Turbulence Spectrum

Fernanda L. dos Santos*, Laura Botero-Bolívar †, Cornelis H. Venner ‡ and Leandro D. de Santana, §
*University of Twente, Drienerlolaan 5, 7522 NB, Netherlands*

**Noise pollution caused by inflow turbulence is a major problem in many applications, such as propellers and fans. Leading edge noise models, e.g., Amiet's model, are widely applied to predict the noise produced by these applications. This noise prediction model relies on the accuracy of the turbulence spectrum, which is usually assumed to be the von Kármán energy spectrum for isotropic turbulence. However, the von Kármán spectrum does not model accurately the dissipation range of the turbulent energy, resulting in incorrect far-field noise predictions for the high-frequency range. An exponential correction can be applied to the spectrum to model the dissipation range. This correction depends on the dissipation frequency, which is the frequency where the energy spectrum changes from a dependence of $k^{-5/3}$ to an exponential decay. This study experimentally investigates nearly isotropic inflow turbulence and determines the flow field characteristics that affect the dissipation frequency in order to model this frequency and the dissipation range. Experiments have been conducted with two passive grids and hot-wire anemometry in the Aeroacoustic Wind Tunnel of the University of Twente. The turbulence uniformity in the test section and the turbulence development in the streamwise direction were analyzed, showing that the grid generated turbulence was mostly uniform and nearly isotropic. The dissipation frequency was observed to depend on the turbulence intensity, the free-stream velocity, and the turbulence length scale. An expression to compute this frequency is proposed, as well as a formula to predict the dissipation range. The predicted leading edge noise is affected by the dissipation range modelling in the high-frequency range, presenting a decrease in level up to 17 dB for the highest frequencies.**

## I. Nomenclature

| | | |
|---|---|---|
| $a_f, b_f, c_f$ | = | fitting constants [-] |
| $b$ | = | grid bar width [m] |
| $B$ | = | constant of the function $f_\eta$ [-] |
| $c$ | = | half-chord length [m] |
| $c_o$ | = | speed of sound [m/s] |
| $d$ | = | half-span length [m] |
| $E(k)$ | = | energy spectrum [m$^3$/s$^2$] |
| $f$ | = | frequency [Hz] |
| $f_d$ | = | dissipation frequency [Hz] |
| $f_k$ | = | Kolmogorov frequency [Hz] |
| $k$ | = | wavenumber [1/m] |
| $k_e$ | = | average wavenumber of the energy-containing eddies [1/m] |
| $k_1, k_x$ | = | wavenumber in the *x*-direction [1/m] |
| $K_x$ | = | wavenumber in the *x*-direction ($K_x = \omega/U$) [1/m] |
| $k_d$ | = | dissipation wavenumber [1/m] |
| $k_y$ | = | wavenumber in the *y*-direction [1/m] |
| $k_z$ | = | wavenumber in the *z*-direction [1/m] |
| $\mathscr{L}$ | = | aeroacoustic transfer function [-] |

*PhD candidate, Engineering Fluid Dynamics, Department of Thermal Fluid Engineering, f.l.dossantos@utwente.nl
†PhD candidate, Engineering Fluid Dynamics, Department of Thermal Fluid Engineering, l.boterobolivar@utwente.nl
‡Chair, Engineering Fluid Dynamics, Department of Thermal Fluid Engineering, AIAA member, c.h.venner@utwente.nl
§Assistant Professor, Engineering Fluid Dynamics, Department of Thermal Fluid Engineering, AIAA member, leandro.desantana@utwente.nl

| | | |
|---|---|---|
| $l_y$ | = | spanwise correlation length [m] |
| $M$ | = | grid mesh size [m] |
| $n$ | = | exponent of the function $f_\eta$ [-] |
| $N$ | = | number of samples [-] |
| $P_{i,j}$ | = | cross-power spectral density [Pa$^2$/Hz] |
| $P_{i,i}$ | = | auto-power spectral density [Pa$^2$/Hz] |
| $R^2$ | = | coefficient of determination [-] |
| $S_{pp}$ | = | far-field power spectral density [Pa$^2$/Hz] |
| $Tu$ | = | turbulence intensity in the $x$-direction [-] |
| $Tw$ | = | turbulence intensity in the $z$-direction [-] |
| $U$ | = | free-stream velocity [m/s] |
| $u$ | = | velocity fluctuation in the $x$-direction [m/s] |
| $\overline{u}$ | = | mean velocity in the $x$-direction [m/s] |
| $u_{rms}$ | = | root-mean-square of the velocity fluctuation in the $x$-direction [m/s] |
| $w$ | = | velocity fluctuation in the $z$-direction [m/s] |
| $\overline{w}$ | = | mean velocity in the $z$-direction [m/s] |
| $x$ | = | downstream direction [m] |
| $x_{LE}$ | = | $x$ position where the airfoil leading edge is located [m] |
| $x_o, y_o, z_o$ | = | observer position [m] |
| $y$ | = | wind tunnel height or spanwise direction [m] |
| $z$ | = | wind tunnel width or normal-to-the-surface direction [m] |
| $\beta$ | = | grid porosity [-] |
| $\gamma^2$ | = | coherence [-] |
| $\Gamma$ | = | gamma function [-] |
| $\Delta y$ | = | mesh size for the hot-wire measurements in the y-direction [m] |
| $\Delta z$ | = | mesh size for the hot-wire measurements in the $z$-direction [m] |
| $\epsilon$ | = | turbulent kinetic energy dissipation rate [m$^2$/s$^3$] |
| $\eta$ | = | Kolmogorov length scale [m] |
| $\eta_y$ | = | y-direction distance between the hot-wire probes [m] |
| $\lambda_f$ | = | longitudinal Taylor microscale [m] |
| $\Lambda_f$ | = | integral length scale [m] |
| $\Lambda_g$ | = | lateral length scale [m] |
| $\nu$ | = | fluid kinematic viscosity [m$^2$/s] |
| $\rho$ | = | fluid density [kg/m$^3$] |
| $\Phi_{uu}, E_{11}(k)$ | = | power spectrum density of the velocity in the $x$-direction [(m/s)$^2$/Hz] |
| $\Phi_{ww}, E_{22}(k)$ | = | power spectrum density of the velocity in the $z$-direction [(m/s)$^2$/Hz] |
| $\omega$ | = | angular frequency ($\omega = 2\pi f$) [1/s] |

## II. Introduction

Flow-induced noise impacts people's well–being and animal life, and urgently needs to be reduced. The societal relevance of aerodynamic noise has contributed to the intensification of this research topic [1, 2]. Leading edge noise is generated by the interaction of a region of unsteady flow or turbulence with a foil or other fluid dynamic devices [3]. This noise source is dominant for applications that are exposed to a turbulent inflow [4, 5], such as propellers [6] and fans [7]. Other sources of broadband noise are also present, such as the noise produced by the interaction between a turbulent boundary layer and the trailing edge (TE) of a foil (trailing edge noise) [3, 7]. These other contributions still exist in the absence of an inflow turbulence, but they become a secondary noise source when the inflow turbulence is at least stronger than 2 or 3% of the mean flow [4, 5, 7]. Therefore, for these applications, the understanding and modelling of the leading edge noise is paramount.

Leading edge noise prediction models are widely used in the design process of propellers, fans, etc., aiming to design more silent equipment. Amiet's leading edge noise prediction model [8] is one of the most applied models due to the simplicity and short turnaround time. This model requires as input flow parameters (e.g., free-stream velocity), geometric parameters (e.g., chord and span length), and inflow turbulence spectrum. Therefore, it is essential to accurately model the turbulent inflow reaching the lifting surface.

The von Kármán energy spectrum [9] is the most commonly used turbulence spectrum model. Von Kármán [9] proposed an energy spectrum model for isotropic turbulence that considers the Kolmogorov dependence in the inertial subrange ($k^{-5/3}$) and the dependence for small wavenumbers ($k^4$) [10]. This modelling has been widely applied and is considered satisfactory in various areas. It gives good approximations for complex terrains in wind farms [11], grid generated turbulence in wind tunnels [12–14], and it has been extensively applied in leading edge noise prediction models [8, 12–14]. Though, for grid generated turbulence, the model does deviate from experimental data in the high-frequency range, i.e., the dissipation range, as observed in [3, 12, 13, 15]. This deviation occurs because in reality the spectrum decays more rapidly than a power of the wavenumber $k$ in the dissipation range [15]. For this reason, the von Kármán spectrum model has to be corrected.

Pope [15] and Roger [13] proposed corrections to the von Kármán spectrum in order to model the dissipation range. Pope [15] suggested that the von Kármán energy spectrum $E(k)$ should be multiplied by the factor $e^{-2.1k\eta}$, where $\eta$ is the Kolmogorov length scale. Roger [13] proposed a similar expression for the correction factor: $e^{-(9/4)(f/f_k)^2}$. Instead of defining the correction factor as a function of the wavenumber and the Kolmogorov length scale, Roger [13] defined the factor with related variables: the frequency ($f$) and the Kolmogorov frequency ($f_k = U/\eta$). However, Pope [15] and Roger [13] did not elaborate on how to determine the Kolmogorov frequency or length scale. Ting [16] presented a formulation for the Taylor microscale ($\lambda_f$), which is directly related to $\eta$ (see section III.F), as a function of the free-stream velocity, the grid mesh, and the distance from the grid. These parameters are strictly applied to grid generated turbulence only but the dissipation range should not depend on how the turbulence is generated since the small scales present a universal behavior. Thus, an expression to the determine the Kolmogorov frequency/length scale from general parameters that characterize the flow and the turbulence is lacking in the literature. The determination of this frequency/length scale is paramount since it dictates where the dissipation range (high frequencies/wavenumbers) begins, i.e., where the energy spectrum changes from a dependence of $k^{-5/3}$ to an exponential decay, as can be seen in Fig. 1. In this study, the frequency where the dissipation range starts is referred as dissipation frequency. The aim of this paper is to experimentally study nearly isotropic turbulent inflows and investigate the flow field characteristics that affect the dissipation frequency (consequently the Kolmogorov length scale) in order to model the dissipation frequency and the spectrum dissipation range.

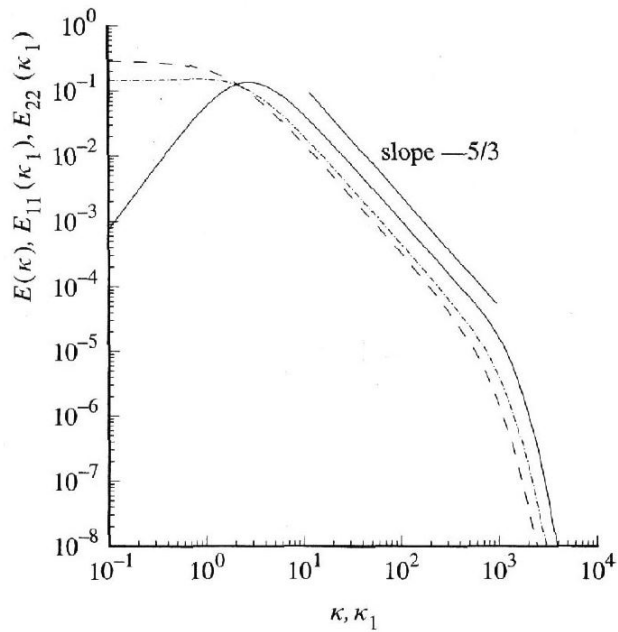

**Fig. 1  Comparison of spectra in isotropic turbulence: solid line, energy spectrum function $E(k)$; dashed line, one-dimensional energy spectrum in the streamwise direction $E_{11}(k_1)$; dashed-dot line, one-dimensional energy spectrum in a streamwise perpendicular direction $E_{22}(k_1)$. $k$ is the wavenumber and $k_1$ is the streamwise wavenumber. Source: Pope [15].**

Grids are the most common technique applied in wind tunnel tests to generate a turbulent inflow. Grids are classified in two types: active and passive. Active grids generate turbulence by adding momentum into the flow by air jets or moving parts [17]. Due to the injection of momentum into the flow field, a high level of turbulence can be achieved, with a lower pressure drop, which results in higher flow velocities downstream of the grid. However, designing and building active grids is considerably more complex, and setting the grid parameters to obtain the desired turbulence characteristics is time consuming [14]. Passive grids block the flow field and generate turbulence with the energy available in the flow. This type of grid achieves lower levels of turbulence and velocities and requires multiple grids to generate different inflow turbulence characteristics. However, they are easier to design and construct. In this research, passive grids were used to generate isotropic turbulent inflow.

The objective of this study is to experimentally investigate nearly isotropic inflow turbulence and determine the flow field characteristics that affect the dissipation frequency in order to model this frequency and the dissipation range decay. To accomplish that, two passive grids were manufactured and hot-wire measurements were performed in the UTwente AeroAcoustic Wind Tunnel. Both grids were designed to generate a nearly isotropic turbulence at $x/M = 10$ (airfoil leading edge position), which was confirmed by the measurements. We determined the flow uniformity, the turbulence intensity decay, the length scale development, and the turbulence spectrum with the downstream position for different free-stream velocities. The measurements were performed at downstream positions ranging from $x/M = 6.5$ to $x/M = 13$, and at free-stream velocities from 10 to 30 m/s. Grid 1 and 2 resulted in turbulence intensities of around 20% and 10% at $x/M = 10$. It was observed that the dissipation frequency depends on the turbulence intensity, the free-stream velocity, and the turbulence integral length scale. A relation to compute this frequency is proposed together with an expression to model the spectrum decay in the dissipation range.

The first section of the paper describes the experimental setup, the procedure applied to design the grids, the measurements performed, and the analyses carried out. The following section discusses the main results obtained from the measurements. Starting by analysing the flow uniformity in the test section, and followed by the turbulence intensity and integral length scale development with the downstream position. The experimental turbulence spectrum obtained from the velocity measurements is compared with the von Kármán model for different downstream positions. The dependence of the dissipation frequency on the flow field parameters is discussed and an expression to compute this frequency is proposed, as well as an expression to model the dissipation range decay. The influence of the dissipation range modelling in the leading edge noise prediction is addressed. Finally, the last section states the main conclusions of the study.

## III. Experimental setup and methodology

### A. Wind tunnel

The experiments were performed in the Aeroacoustic Wind Tunnel of the University of Twente which is an aeroacoustic open-jet closed-circuit wind tunnel. The test section dimensions are 0.7 m × 0.9 m and it has a contraction ratio of 10:1. After the contraction, the flow enters a closed test section (CTS) and subsequently an open test section (OTS). The maximum operating velocity is 60 m/s in open-jet configuration with a turbulence intensity below 0.08% [18]. The flow temperature was controlled and maintained at approximately 20 °C for all experiments. The grids were installed in the CTS and hot-wire measurements were performed in the open test section (see Fig. 2a). The grid position in the CTS was chosen to guarantee that the grid is at least 10 times the mesh size upstream from the leading edge (LE) of any airfoil installed in the OTS ($x_{LE}/M > 10$). This assures that the inflow turbulence that interacts with the airfoil leading edge is homogeneous and fully developed, according to Roach [19]. The coordinate reference system considers the downstream-direction axis $x$, the spanwise-direction axis $y$, and the normal-to-the-surface direction axis $z$ with the origin at the grid position and at the test section center.

### B. Turbulence grids

Two passive grids with squared bars were designed (Fig. 3). The first grid (grid 1) was projected to generate the highest level of turbulence and the longest length scale possible with porosity higher than 30%. The second grid (grid 2) was projected to generate an intermediate level of turbulence (around 7%) with porosity higher than 50%. Both grids satisfied the relation $x_{LE}/M > 10$ proposed by Roach [19] to generate an isotropic turbulence at the airfoil leading edge plane in the OTS. For simplicity, both grids were installed at the same location, which lead to a constant $x_{LE} = 1.4$ m. The grid bar width was designed based on a trade-off of the turbulence level, the length scale, and the porosity desired.

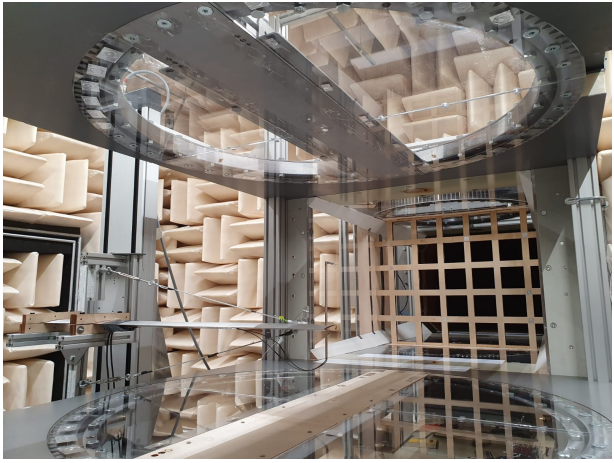
(a) Experimental setup.

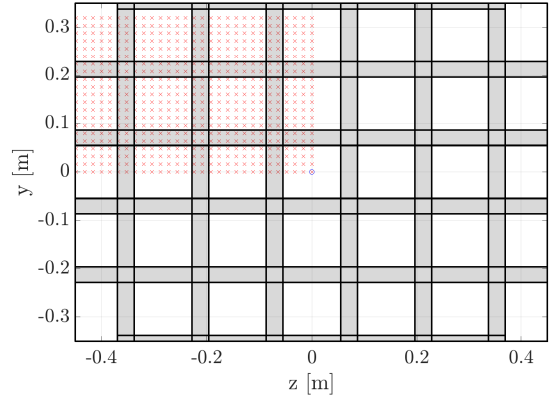
(b) Hot-wire measurement positions.

Fig. 2   a) Experimental setup: hot-wire support located at the wind tunnel open test section and grid 2 installed at the closed test section. b) Hot-wire measurement points overlapping with the bars of grid 2 in the upper left quadrant of the test section. Red x's indicate the positions where the velocity mapping was performed (see section III.C.1). Blue circle indicate the position where turbulence development was determined (see section III.C.2).

Table 1   Grid projected parameters. Predicted values at the leading edge plane determined by Eqs. 1– 4.

|  | b [m] | M [m] | $\beta$ | Tu [%] | Tw [%] | $\Lambda_f$ [mm] |
|---|---|---|---|---|---|---|
|  |  |  |  | Predicted at the LE plane | | |
| Grid 1 | 0.06 | 0.14 | 0.3 | 12.4 | 11.0 | 68 |
| Grid 2 | 0.03 | 0.14 | 0.6 | 7.5 | 6.7 | 48 |

Roach [19] introduced simple equations to describe the turbulence decay (Eqs. 1 and 2) and length scale development (Eq. 3) with the downstream direction ($x$) based on the grid bar width ($b$). These equations together with the porosity definition (Eq. 4) and the restrictions were used to determine the grid bar width. Table 1 shows the grid geometric parameters and the predicted turbulence intensities and length scales at the leading edge plane.

$$Tu = 1.13 \cdot \left(\frac{x}{b}\right)^{-5/7} , \quad (1)$$

$$Tw = 0.89 \cdot 1.13 \cdot \left(\frac{x}{b}\right)^{-5/7} , \quad (2)$$

$$L = 0.2 \cdot b \cdot \left(\frac{x}{b}\right)^{1/2} , \quad (3)$$

$$\beta = \left(1 - \frac{b}{M}\right)^2 . \quad (4)$$

The maximum mean free-stream velocity achieved with the grids installed in the wind tunnel was not the same due to the different pressure drop introduced by each grid. Grid 1 had a considerably smaller porosity which caused a significant pressure drop. This limited the maximum velocity at the leading edge plane to approximately 10 m/s. Grid 2 had a larger open area, which resulted in a smaller pressure drop compared to grid 1. Grid 2 achieved a maximum free-stream velocity of 30 m/s. The conditions at which measurements were performed are summarized in Table 2.

## C. Hot-wire measurements

Hot-wire measurements were performed to determine the flow characteristics generated by the grids. An X hot-wire probe (55P51 - Dantec Dynamics) of 5 μm diameter and 3 mm wire length was used to measure the *x*- and *z*-components of the velocity. The probe was attached to the Dantec probe support 22H26, which was installed in a symmetric airfoil (see Fig. 2a). The airfoil was fixed in the Dantec 3D traverse system, which was responsible for the probe translation. This system has a resolution of 6.5 μm. The hot-wire velocity calibration was carried out in-situ. The velocity was calibrated using a Prandtl tube as a reference, varying from 0 m/s to 60 m/s with 21 points distributed logarithmically, including a calibration point at 0 m/s. The directional calibration was performed using the THORLABS motorized rotation mount model K10CR1A2/M (absolute accuracy of ±0.14°) in the range ±45° with an interval of 5°. The data was acquired by the Dantec StreamLine Pro CTA system and the Dantec StreamWare software in combination with the National Instruments 9215 A/D converter. From the square wave test, the hot wire had a frequency response exceeding 75 kHz. Ashok et al. [20] used a similar hot-wire probe and similar flow conditions as in this study and they showed that to avoid temporal filtering (inadequate frequency response) the hot-wire frequency response should be at least 37.5 kHz. Therefore, temporal filtering is negligible for the hot-wire measurements.

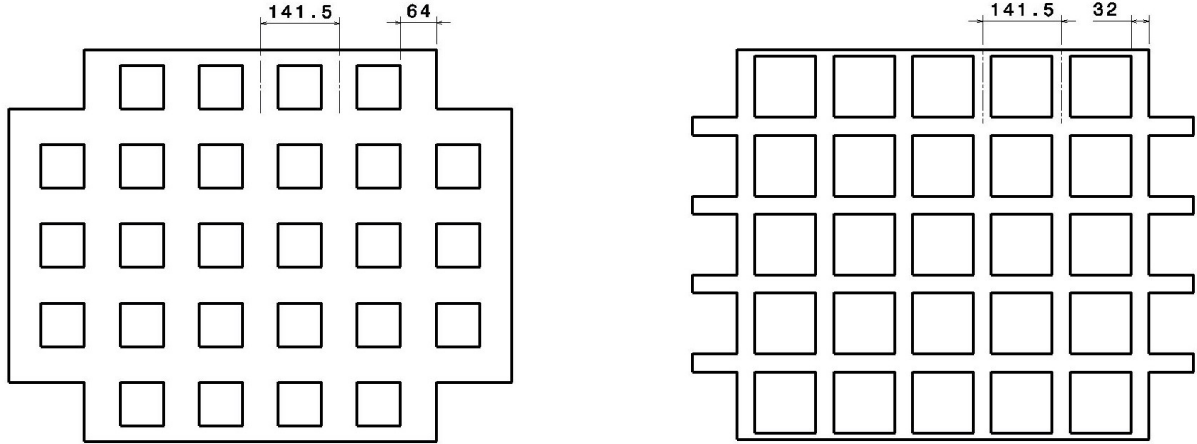

**Fig. 3  Manufactured grids: grid 1 (left) and grid 2 (right). Dimensions are in mm.**

**Table 2  Measurement conditions.**

|  | Velocity mapping | | | | Turbulence development | | | |
|---|---|---|---|---|---|---|---|---|
|  | U [m/s] | x [m] | Δy [mm] | Δz [mm] | U [m/s] | x [m] | y [mm] | z [mm] |
| No grid | 10, 30 | 0.9, 1.4 | 16 | 16 | - | - | - | - |
| Grid 1 | 10 | 0.9, 1.4 | 16 | 16 | 10 | 0.9-1.8 (Δ = 0.1) | 0 | 0 |
| Grid 2 | 10, 30 | 0.9, 1.4 | 16 | 16 | 10:5:30 | 0.9-1.8 (Δ = 0.1) | 0 | 0 |

|  | Lateral length scale | | | |
|---|---|---|---|---|
|  | U [m/s] | x [mm] | Δy [mm] | z [mm] |
| Grid 2 | 10:5:30 | 1.4 ($x_{LE}$) | 0.5 | 0 |

The velocity measurements allowed the determination of the uniformity in the test section, the integral length scale, the turbulence intensity, the turbulence spectrum, and the lateral length scale of the flow generated by the grids. Different measurements were carried out in order to evaluate these parameters and each of them is explained in detail in the following sections.

*1. Velocity mapping*

Velocity mappings were performed to determine the flow uniformity in the test section. The X-wire probe measured the $x$- and $z$-components of the velocity in the upper left quadrant of the open test section (see Fig. 2). The flow was assumed to be symmetric in the test section which was confirmed by measuring the velocities in the lower left quadrant. The velocity mappings were measured in two $y - z$ planes: $x = 0.9$ m (end of CTS), and $x = x_{LE} = 1.4$ m (LE position). It had a spacing of $\Delta z = \Delta y = 16$ mm, which resulted in a total of 609 points per plane (see Fig. 2b and Table 2). The hot-wire data was recorded for 10 s at a sampling frequency of 20 kHz with a low-pass filter of 10 kHz.

*2. Turbulence development*

The turbulence development refers to the evolution of the turbulence intensity ($Tu$ and $Tw$), the integral length scale ($\Lambda_f$) and the turbulence spectrum in the streamwise direction. To determine these parameters the X-wire probe measured the $x$- and $z$-velocity components in the center of the test section (see Fig. 2b and Table 2). The downstream positions ranged from $x = 0.9$ m (end of CTS) to $x = 1.8$ m with an interval of 0.1 m. The acquisition time per point was 30 s because spectral analyses were required. The hot-wire data was acquired at a sampling frequency of approximately 65.5 kHz ($2^{16}$Hz) with a low- and high-pass filter with cut-off frequencies of 30 kHz and 5 Hz, respectively.

*3. Lateral length scale*

The determination of the lateral length scale ($\Lambda_g$) needed the measurement of a 2-point correlation, requiring the use of two X-wire probes in order to quantify the cross-correlation of the $z$-velocity in the $y$-direction. This length scale was measured only for grid 2 because grid 1 could be tested only at 10 m/s. The lateral length scale was evaluated by measuring the $z$-velocity at the LE position ($z = 0$ m and $x_{LE} = 1.4$ m) in different $y$ positions (see Table 2). One X-wire probe was fixed in a $y$ position and the other X-wire probe moved in the $y$-positive direction with an interval of 0.5 mm. For an isotropic turbulence, the lateral length scale is half of the integral length scale [15]. Based on the projected parameters for grid 2, the lateral length scale was expected to be equal to $\Lambda_g \approx 25$ mm. In this way, the maximum distance between the probes was chosen to be three times the expected lateral length scale ($3 \times \Lambda_g \approx 75$ mm). Hence, a total of 151 points were measured with a maximum distance between the probes of 75 mm and an interval of 0.5 mm. The acquisition time per point was 30 s. The hot-wire data was recorded at a sampling frequency of 32.8 kHz ($2^{15}$Hz) with a low- and high-pass filter with cut-off frequencies of 10 kHz and 5 Hz, respectively.

**D. Turbulence characterization from hot-wire data**

*1. Turbulence intensity*

The turbulence intensity was computed following Hinze [21]:

$$Tu = \frac{1}{\bar{u}}\sqrt{\frac{\sum_{i=1}^{N} u^2}{N}}, \qquad (5) \qquad Tw = \frac{1}{\bar{u}}\sqrt{\frac{\sum_{i=1}^{N} w^2}{N}}, \qquad (6)$$

where $u$ and $w$ are the velocity fluctuations in the downstream and the $z$-direction at each measurement point, and $N$ is the number of samples.

*2. Integral length scale*

The integral length scale ($\Lambda_f$), also known as the longitudinal length scale, was computed based on the methodology proposed by Hinze [21]. Firstly, the autocorrelation of the downstream velocity time series is computed. Subsequently, the time when the autocorrelation reaches zero for the first time is considered as the turbulence time scale (see Fig. 4). In the light of Taylor's hypothesis of frozen turbulence, the integral length scale is computed from the time scale considering that the turbulence is being convected with the mean velocity of the time series. This methodology considering Taylor's hypothesis was chosen because the experimental setup needed is relatively simple requiring only one probe to acquire the time series instead of using two probes with varying spacing between them, and it leads to accurate results [15].

*3. Lateral length scale*

The lateral length scale ($\Lambda_g$) is the length scale of the largest eddies in the $z$-direction in a turbulent flow. The lateral length scale is an important parameter for the inflow characterization because its comparison with the integral length scale indicates the isotropy of the flow. If a turbulent flow is isotropic, the lateral length scale should be half of the integral length scale ($\Lambda_g = \Lambda_f/2$). To compute this scale, first the coherence ($\gamma^2$) of the $z$-velocity component between the X-wire probes was calculated for each spacing between the probes ($\eta_y$). The coherence is a normalized quantity computed from the auto- ($P_{i,i}(f)$) and cross-power ($P_{i,j}(f)$) spectral density of the $z$-velocity data signal, defined as [22]:

$$\gamma^2(f) = \frac{|P_{i,j}|^2}{P_{i,i}(f) P_{j,j}(f)}. \qquad (7)$$

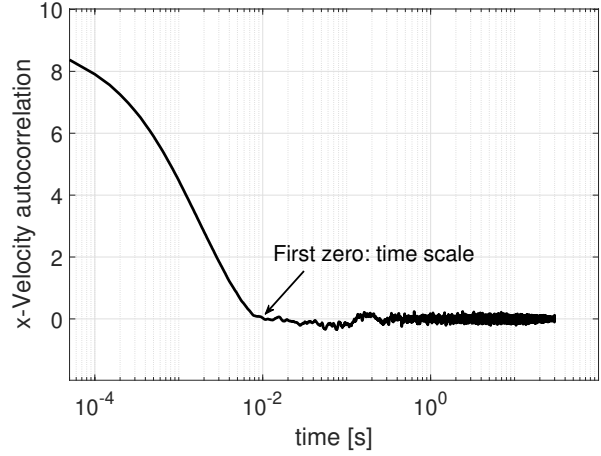

**Fig. 4** Autocorrelation of the streamwise velocity for grid 2 for a free-stream velocity of 30 m/s. The first time that the autocorrelation reaches zero is indicated and this point is considered as the turbulence time scale, as proposed by Hinze [21].

In this way, for each spacing between the X-wire probes ($\eta_y$), the coherence was determined as a function of the frequency. From the coherence, the spanwise correlation length is defined as [4, 8, 23]:

$$l_y(f) = \int_0^\infty \sqrt{\gamma^2(f,\eta_y)}\,\mathrm{d}\eta_y\,. \tag{8}$$

From this equation, the spanwise correlation length is determined for different frequencies. The lateral length scale is defined as the length scale of the biggest eddies, corresponding to the spanwise correlation length for low frequency.

*4. Turbulence spectrum*

The turbulence spectra $\Phi_{uu}$ and $\Phi_{ww}$ are the power spectral density of the velocities in $x$- and $z$-direction, respectively. The spectrum is evaluated from the velocity time series based on Welch's method. The data was averaged using blocks of $2^{16} = 65536$ (1 s) samples and windowed by a Hanning windowing function with 50% overlap, resulting in a frequency resolution of 1 Hz. The turbulence spectrum is generally defined in terms of the wavenumber in the downstream direction ($k_x$), making the following relations useful [12]:

$$\Phi_{uu}(f) = \frac{\bar{u}}{2\pi}\Phi_{uu}(k_x)\,, \tag{9}$$

$$\Phi_{ww}(f) = \frac{\bar{u}}{2\pi}\Phi_{ww}(k_x)\,. \tag{10}$$

The downstream $\Phi_{uu}(k_x)$ and the $z$-direction $\Phi_{ww}(k_x)$ turbulence spectra calculated from the experimental data were compared with the von Kármán spectrum. The von Kármán energy spectrum function is given as [3, 24]:

$$E(k) = \frac{55}{9\sqrt{\pi}}\frac{\Gamma(5/6)}{\Gamma(1/3)}\frac{\overline{u^2}}{k_e}\frac{(k/k_e)^4}{\left[1+(k/k_e)^2\right]^{17/6}}\,, \tag{11}$$

where $k_e$ is the wavenumber scale of the largest eddies [3], defined as [3, 12]:

$$k_e = \frac{\sqrt{\pi}}{\Lambda_f}\frac{\Gamma(5/6)}{\Gamma(1/3)}\,. \tag{12}$$

Equation 11 can be integrated into other spectral forms that are of fundamental interest to determine the turbulence characteristics. The spectral forms that are the most useful for this study are: $\Phi_{ww}(k_x,k_y)$, $\Phi_{uu}(k_x)$, and $\Phi_{ww}(k_x)$.

The von Kármán planar turbulence spectrum is obtained by integrating the energy spectrum (Eq. 11) with respect to $k_z$ [3, 8, 24]:

$$\Phi_{ww}|_{vK}(k_x, k_y) = \int_{-\infty}^{\infty} \frac{E(k)}{4\pi k^2}\left(1 - \frac{k_z^2}{k^2}\right) dk_z = \frac{4}{9\pi} \frac{\overline{u^2}}{k_e^2} \frac{(k_x/k_e)^2 + (k_y/k_e)^2}{[1 + (k_x/k_e)^2 + (k_y/k_e)^2]^{7/3}}. \quad (13)$$

The von Kármán one-dimensional spectra are given as [12]:

$$\Phi_{uu}|_{vK}(k_x) = \int_{-\infty}^{\infty}\int_{-\infty}^{\infty} \frac{E(k)}{4\pi k^2}\left(1 - \frac{k_x^2}{k^2}\right) dk_z dk_y = \frac{2}{\sqrt{\pi}} \frac{\Gamma(5/6)}{\Gamma(1/3)} \frac{\overline{u^2}}{k_e} \left[1 + \left(\frac{k_x}{k_e}\right)^2\right]^{-5/6}, \quad (14)$$

$$\Phi_{ww}|_{vK}(k_x) = \int_{-\infty}^{\infty}\int_{-\infty}^{\infty} \frac{E(k)}{4\pi k^2}\left(1 - \frac{k_z^2}{k^2}\right) dk_z dk_y = \frac{1}{2\sqrt{\pi}} \frac{\Gamma(5/6)}{\Gamma(1/3)} \frac{\overline{u^2}}{k_e} \frac{1 + (8/3)(k_x/k_e)^2}{(1 + (k_x/k_e)^2)^{11/6}}. \quad (15)$$

### E. Leading edge noise prediction model

Amiet [8] proposed a semi-analytical model to predict the far-field noise of an airfoil in a turbulent stream. The far-field noise is calculated from the inflow turbulence spectrum ($\Phi_{ww}(k_x, k_y)$), usually modelled by the von Kármán turbulence spectrum. The model assumes an infinitely large span, a flat plate geometry with an infinitely small thickness, a stationary observer, and a uniform free-stream velocity. Equation 16 presents the far-field power spectral density of a flat plate of chord $2 \times c$ and span $2 \times d$ considering an observer in the plane $y = 0$. The coordinate reference system is the same as discussed in section III.A but for the noise prediction model the origin is located in the mid-chord and mid-span of the airfoil surface.

$$S_{pp}(x_o, y_o = 0, z_o, \omega) = \left(\frac{\omega z_o \rho c}{c_o \sigma^2}\right) \pi U d |\mathscr{L}(x, K_x, 0)|^2 \Phi_{ww}(K_x, 0). \quad (16)$$

The aeroacoustic transfer function $\mathscr{L}$ is determined from de Santana [12] (pp.155, 166 and 168). The von Kármán turbulence spectrum for Amiet's model ($\Phi_{ww}|_{vK}$) is a function of $k_x$ and $k_y$, and it is given by Eq. 13.

### F. Modelling of the spectrum dissipation range

The von Kármán turbulence spectrum has a dependence of $k^{-5/3}$ in the inertial subrange, as can be seen in Fig. 1. However, the exponential decay for high wavenumbers (dissipation range) is not modelled. Pope [15] extended the von Kármán model to include the dissipation range (see Fig. 1). Pope's dissipation range model involves the multiplication of the energy spectrum Eq. 11 by the function:

$$f_\eta|_{\text{Pope}}(k, \eta) = \exp(-B(k\eta)^n), \quad (17)$$

where $B \approx 2.1$, $n = 1$ and $\eta$ is the Kolmogorov length scale. Roger [13] proposed a similar equation as Eq. 17 with $B = 9/4$ and $n = 2$. The Kolmogorov length scale is the smallest length scale in a turbulent flow and is defined as [16]:

$$\eta = \left(\frac{\nu^3}{\epsilon}\right)^{1/4}, \quad (18)$$

where $\epsilon$ is the average rate of dissipation of the turbulence kinetic energy. Assuming isotropic dissipation, the dissipation rate is given as [16]:

$$\epsilon = 15\nu \frac{\overline{u^2}}{(2\lambda_f)^2}, \quad (19)$$

where $\lambda_f$ is the longitudinal Taylor microscale. The Taylor microscale corresponds to the largest eddy length scale in the dissipation range.

In this study, we modeled the frequency where the dissipation range begins ($f_d$) and correlate this frequency with the Taylor microscale by defining $\lambda_f = U/f_d$. The dissipation frequency from the experimental data was determined as the frequency where the slope of the experimental velocity spectrum deviated 50% from the slope of the von Kármán spectrum. The dissipation frequency was modeled as a function of flow and turbulence parameters instead of grid parameters as suggested by Ting [16]. This formulation was chosen because it makes the model more general, i.e., not grid-parameter dependent. This type of formulation is essential because the dissipation range should not depend on how the turbulence is generated since the small scales present a universal behavior. By modeling the dissipation frequency, the Taylor microscale, the dissipation rate, and consequently the Kolmogorov length scale can be calculated. A similar equation to Eq. 17 was applied to model the dissipation range decay with the constant $B$ and $n$ fitted to the cases studied. The results and the empirical relations are presented in section IV.E.

## IV. Results

First, flow uniformity in the test section is discussed, followed by the turbulence intensity and the integral length scale development with the downstream position. Subsequently, the lateral length scale is analysed. Finally, the experimental turbulence spectrum obtained from the velocity measurements is compared with the von Kármán model for different downstream positions, where the modelling of the dissipation frequency and the dissipation range is discussed. Finally, the model for the dissipation range proposed in this study is applied to the Amiet's leading edge noise prediction model.

### A. Flow uniformity

The flow uniformity was verified by analyzing the mean velocities and the turbulence intensities for the cases without grid and with grids, as described in section III.C. The $y$ and $z$ locations where the measurements were performed are depicted in Fig. 2b by red crosses. The flow uniformity without grid was verified, showing that the velocities and turbulence intensities were extremely uniform in the test section. The presence of the shear layer due to the open test section configuration and the boundary layer due to the upper wall was clearly visible from the velocity mappings without grid. The mappings for this condition are not included for sake of brevity.

Figure 5 shows the $x$- and $z$- direction velocities and turbulence intensities for grid 1 at 10 m/s and $x_{LE} = 1.4$ m in the upper left quadrant of the test section (see Fig. 2b). From Fig. 5, one can see that the flow velocities were not fully uniform in the test section. However, the region of interest is in the test section center ($z = 0$ m) because the LE of the airfoil would be located at this position. At this region, the velocities are more uniform and the turbulence intensities are considerably uniform in the $y$ direction. Similar results were obtained for $x = 0.9$ m; however, the velocities at this location were slightly higher. The average turbulence intensities at the center of the test section were $Tu = 20.5\%$ and $Tw = 17.2\%$.

Figure 6 shows the $x$- and $z$- direction velocities and turbulence intensities for grid 2 at 30 m/s and $x_{LE} = 1.4$ m in the upper left quadrant of the test section (see Fig. 2b). For this grid, the $x$-direction velocity was not fully uniform; however, it was more homogeneous than for the grid 1. This is probably because of the higher porosity of grid 2 compared to grid 1, which disturbs the flow considerably less. The $z$-direction velocity and the turbulence intensities were highly uniform for grid 2. Similar results were obtained for 10 m/s and $x = 0.9$ m, with only a slightly difference in the shear layer region. The average turbulence intensities at the center of the test section were $Tu = 10.3\%$ and $Tw = 9.11\%$.

### B. Turbulence intensity development

Velocity measurements were performed at the test section center (blue circles in Fig. 2b) at 10 different downstream positions (check Table 2) to investigate the turbulence intensity development. The turbulence intensity for both the $x$- and $z$- direction velocities were determined at each measurement position.

Figure 7 shows the turbulence intensity development for grids 1 and 2, and the predicted development given by Eqs. 1 and 2. The experimental turbulence intensity roughly followed the same tendency as predicted by Roach [19], except for grid 1 for $x/M > 10$. The reason for this behavior is still unclear. Furthermore, the turbulence intensity did not vary with the free-stream velocity, as observed by Roach [19]. The turbulence intensity values were considerably higher than the ones predicted by Eqs. 1 and 2, with $Tu$ at the LE position 65% higher than predicted ($Tu = 20.5\%$) for grid 1 and 37% higher for grid 2 ($Tu = 10.3\%$). This mismatch is probably because the grids are outside the validity range of the Roach equations, i.e., the grids studied in this paper generated much higher levels of turbulence than the data used by Roach [19] to determine Eqs. 1 and 2.

Due to the poor prediction of the turbulence intensity by the Roach equations, a fitting of these equations to the experimental data was performed to determine the new constants for these equations. The fitted equations are shown in Eqs. 20 and 21 with the values of the constants shown in Table 3 together with the coefficient of determination $R^2$ for each equation fitted. The $R^2$ value for Eq. 21 is unexpected because the fitted curve seems to predict the decay of the turbulence intensity, as can be inferred from Fig 7. The constant $a_f$ fitted for grid 1 is similar to one from Roach [19] (1.13); however, the constant for grid 2 is more than double. This was unexpected because grid 1 generated a much higher level of turbulence, deviating considerably from the predicted values. The exponent constant from Roach [19] (-0.71) is an intermediate value between the $b_f$ values found for both grids. The constant $c_f$ for both grids is extremely close to the value proposed by Roach [19] (0.89).

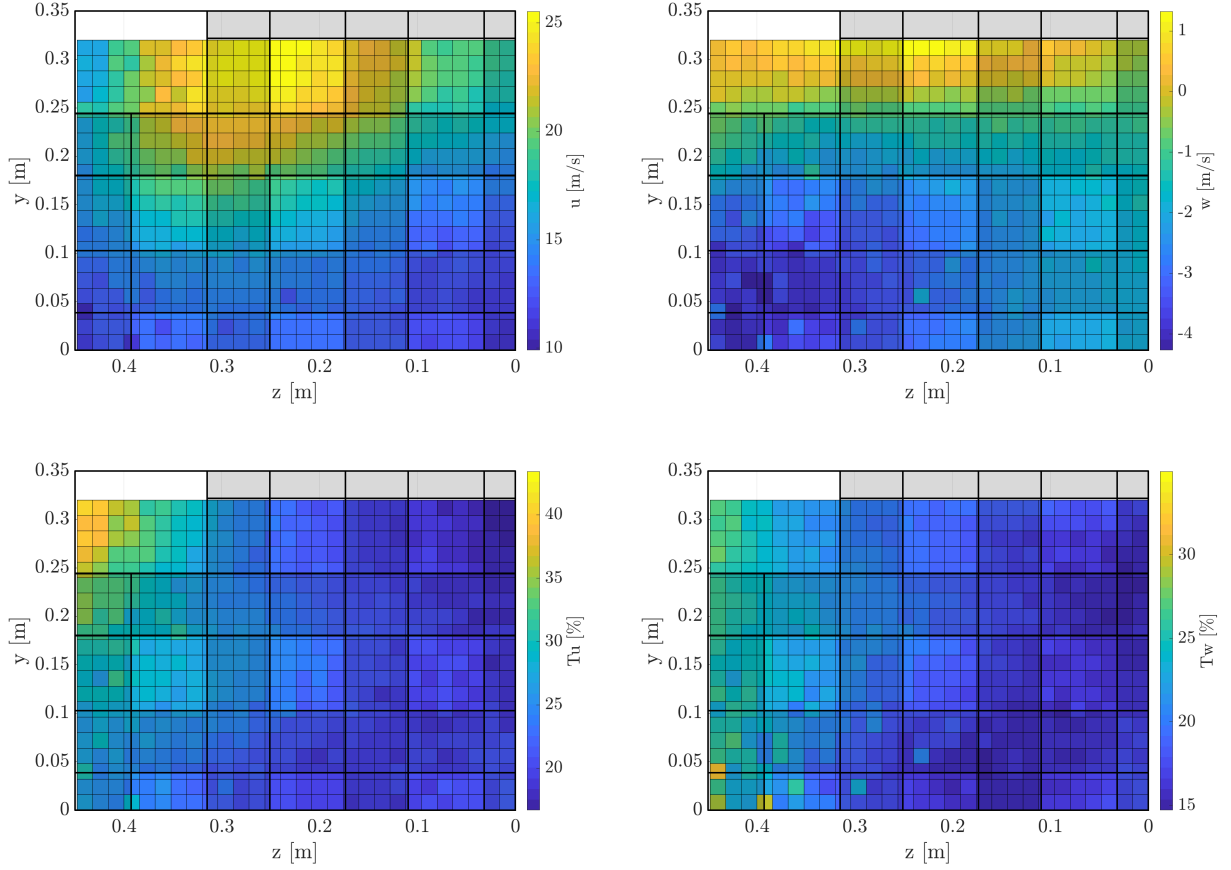

**Fig. 5** Velocity and turbulence intensity mappings for grid 1 at 10 m/s and $x_{\text{LE}}$ = 1.4 m in the upper left quadrant of the test section (see Fig. 2b). The LE position is at $z = 0$ m. The grid bars are shaded. The upper left graph shows the downstream velocity, the upper right the $z$-direction velocity, the lower left the downstream turbulence intensity, and the lower right the $z$-direction turbulence intensity.

$$Tu = a_f \cdot \left(\frac{x}{b}\right)^{b_f}, \qquad (20) \qquad Tw = a_f \cdot c_f \cdot \left(\frac{x}{b}\right)^{b_f}. \qquad (21)$$

**Table 3** Fitting constants and coefficient of determination for $T_u$ and $T_w$ development given by Eqs. 20 and 21.

|  | $a_f$ | $b_f$ | $R^2_{a_f,b_f}$ | $c_f$ | $R^2_{c_f}$ |
|---|---|---|---|---|---|
| Grid 1 | 1.21 | -0.56 | 0.9211 | 0.80 | -0.13 |
| Grid 2 | 3.72 | -0.94 | 0.9903 | 0.88 | 0.9820 |

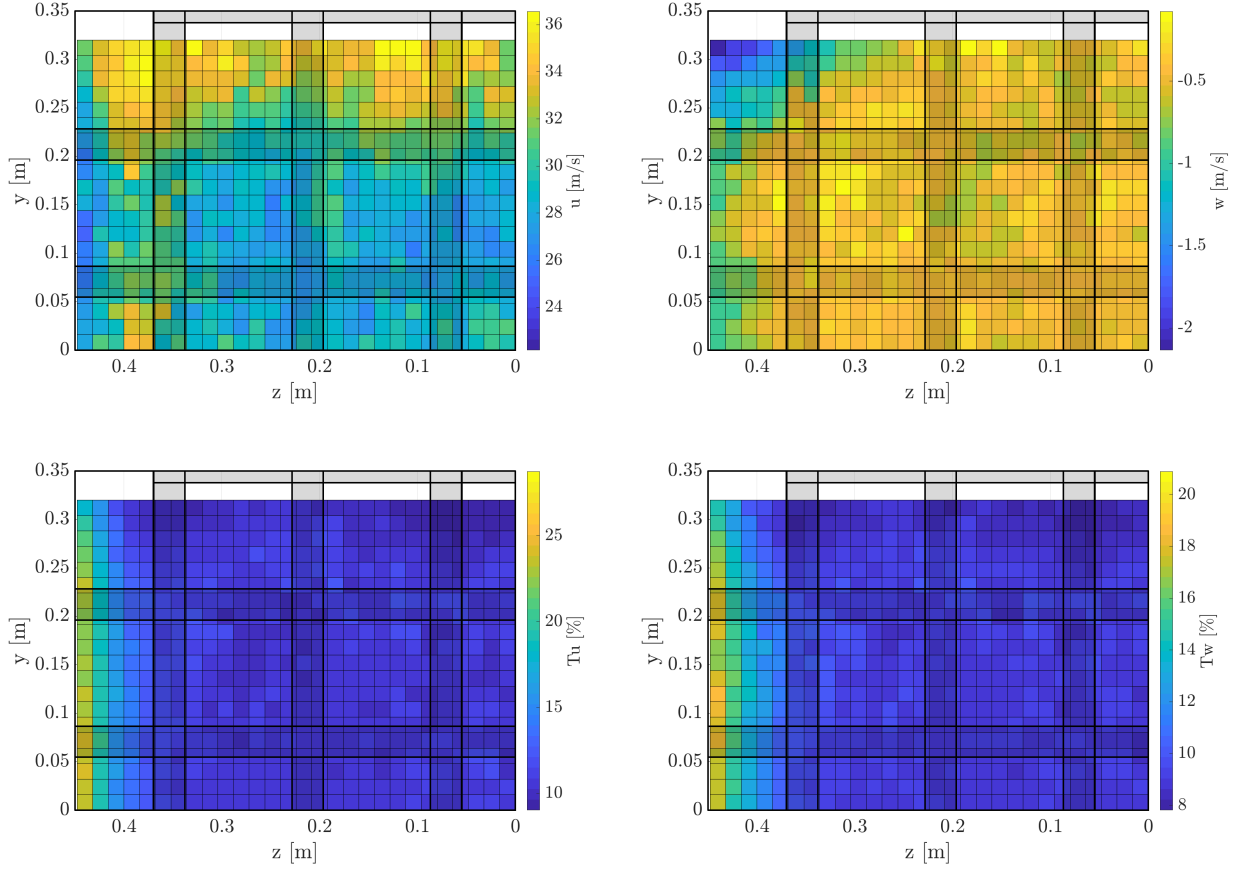

**Fig. 6   Velocity and turbulence intensity mappings for grid 2 at 30 m/s and $x_{LE}$ = 1.4 m in the upper left quadrant of the test section (see Fig. 2b). The LE position is at $z$ = 0 m. The grid bars are shaded. The upper left graph shows the downstream velocity, the upper right the $z$-direction velocity, the lower left the downstream turbulence intensity, and the lower right the $z$-direction turbulence intensity.**

### C. Integral length scale development

To investigate the integral length scale development, velocity measurements were performed at the test section center (blue circles in Fig. 2b) at 10 different downstream positions (check Table 2).

Figure 8 shows the integral length scale development for grids 1 and 2, and the predicted development given by Eq. 3. The experimental integral length scale roughly follows the same tendency as predicted by Roach [19]. However, a remarkable difference is observed: the integral length scale depends on the free-stream velocity. This is probably because of the interaction between the vortex shedding from the different grid bars; however, this behavior has to be further investigated. Due to the velocity dependence, the integral length scale predicted by Eq. 3 is no longer valid.

### D. Lateral length scale

The $z$-velocity spanwise correlation length was computed by Eq. 8, resulting in the correlation length as a function of the frequency. Figure 9 depicts this correlation length for different free-stream velocities measured at the LE position. As expected, the spanwise correlation length decreases as the frequency increases. It reaches a plateau at approximately 10 mm for all velocities. This plateau is believed to be a consequence of the hot-wire spatial filtering due to averaging over the wire length [20]; however, further analysis has to be carried out to confirm this hypotheses. Furthermore, the spanwise correlation length varies with the velocity, presenting the same behavior as observed for the integral length scale.

The lateral length scale is the length of the biggest eddies, corresponding to the low-frequency range of the spanwise

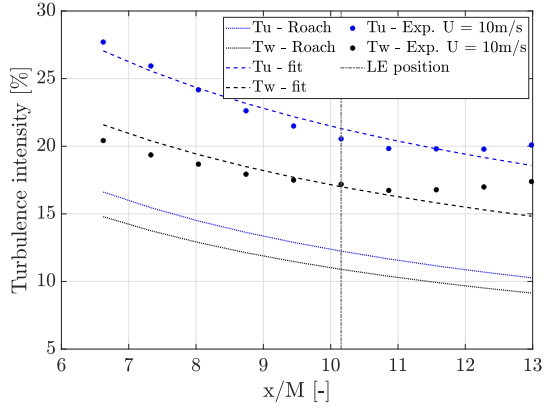
(a) Grid 1.

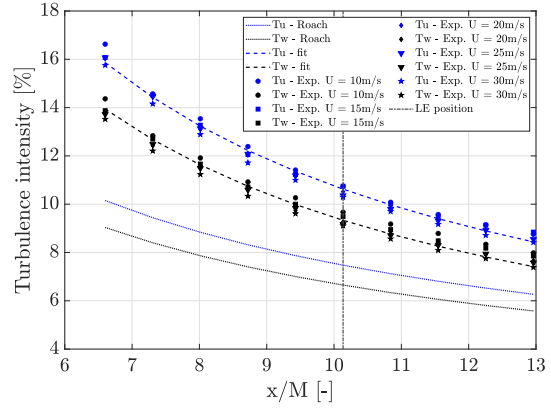
(b) Grid 2.

**Fig. 7** Turbulence intensity development with the downstream position for grid 1 and 2. The curves referred as Roach are given by Eqs. 1 and 2. The curves referred as fit are given by Eqs. 20 and 21. The *x* location where the LE is located is represented by dot-dashed line.

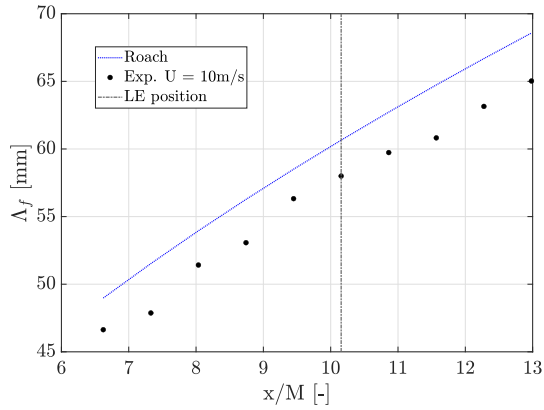
(a) Grid 1

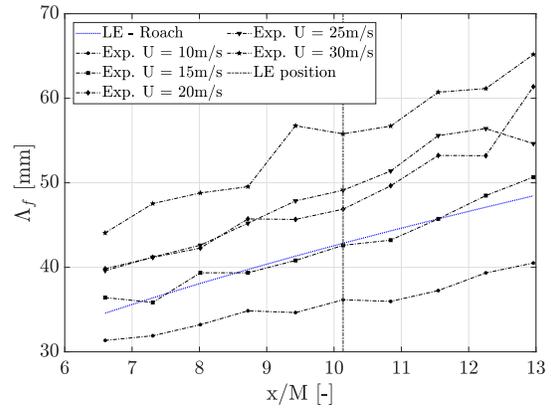
(b) Grid 2.

**Fig. 8** Integral length scale development with the downstream position for grid 1 and 2. The curve referred as Roach is given by Eq. 3. The *x* location where the LE is located is represented by dot-dashed line.

correlation length. Hence, the lateral length scale was considered as the spanwise correlation length at the lowest frequency measured (10 Hz). Table 4 shows the main turbulence parameters at the LE position for the different velocities and grids tested, including the lateral length scale values. For an isotropic turbulence, the lateral length scale is half of the integral length scale [15, 16]. As can be inferred from Table 4, the lateral length scale is roughly half of the integral length scale, indicating that the grid-generated turbulence is nearly isotropic. The lateral length scale was determined based on the lowest frequency from the measurements, which does not mean that it is the lowest frequency describing the phenomenon. However, the same frequency range was considered to compute the integral and lateral length scales making the comparison between them valid.

### E. Turbulence spectrum and dissipation range modelling

The turbulence spectrum was computed for the *x*- and *z*-direction velocities at the test section center (blue circles in Fig. 2b) at different downstream positions (check Table 2). Figures 10 and 11 show the experimental and the von Kármán spectra for grid 1 and 2 for different downstream positions and free-stream velocities. The experimental turbulence spectrum for the streamwise velocity ($\Phi_{uu}|_{\text{exp}}$) matches considerably well with von Kármán spectrum up to the dissipation wavenumber/frequency ($k_d, f_d$) where the dissipation range begins. This shows that the von Kármán model for isotropic turbulence describes the power spectral density of the velocity quite well up to the dissipation range. The experimental turbulence spectrum for the *z*-direction velocity ($\Phi_{ww}|_{\exp}$) deviates slightly from the von Kármán spectrum in the mid-frequency range. The same behavior as presented by $\Phi_{uu}|_{\exp}$ in the dissipation range is also observed for $\Phi_{ww}|_{\exp}$. Depending on the free-stream velocity, the spectrum converges to a plateau, which is due to the hot-wire noise for high frequencies.

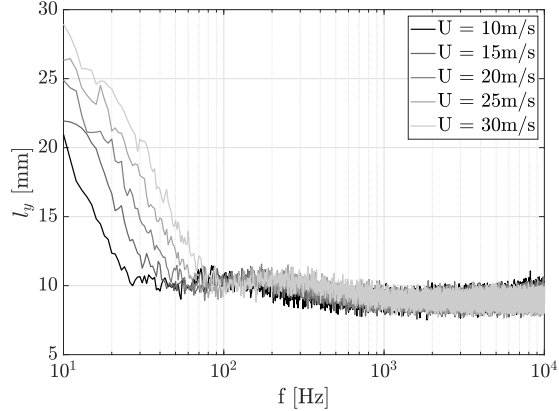

**Fig. 9** Spanwise correlation length scale for the *z*-direction velocity at different free-stream velocities at the LE position.

**Table 4** Turbulence intensities, integral length scale, lateral length scale and ratio between the length scales for grid 1 and 2 at the LE position.

|  | U [m/s] | $Tu$ [%] | $Tw$ [%] | $\Lambda_f$ [mm] | $\Lambda_g$ [mm] | $\Lambda_f/\Lambda_g$ [-] |
|---|---|---|---|---|---|---|
| Grid 1 | 10 | 20.5 | 17.2 | 58 | - | - |
| Grid 2 | 10 | 10.7 | 9.67 | 36 | 21 | 0.58 |
| Grid 2 | 15 | 10.7 | 9.48 | 43 | 22 | 0.51 |
| Grid 2 | 20 | 10.5 | 9.26 | 47 | 25 | 0.53 |
| Grid 2 | 25 | 10.4 | 9.16 | 49 | 26 | 0.53 |
| Grid 2 | 30 | 10.3 | 9.11 | 56 | 29 | 0.52 |

The empirical expression for the dissipation frequency $f_d$ was obtained from the experimental data. The experimental dissipation frequency ($f_d|_{\exp}$) was determined as the frequency where the slope of the experimental velocity spectrum deviated 50% from the slope of the von Kármán spectrum. This frequency is depicted in Figs. 10 and 11 as $k_d|_{\exp}$. It was observed that this frequency is highly dependent on the free-stream velocity and downstream position. It is well-known that the turbulence intensity and the integral length scale vary with the streamwise direction, being directly linked to the downstream position. Hence, an expression to model the dissipation frequency is proposed in the form shown in Eq. 22 as a function of the free-stream velocity, the root-mean-square of the streamwise velocity fluctuations, and the integral length scale. To obtain the constants, the dissipation frequencies determined from the experimental data for $\Phi_{uu}|_{\exp}$ were fitted to the expression shown in Eq. 22. The coefficient of determination for the fitting was $R^2 = 0.98$. The final expression is given as:

$$f_d = 12.18 \cdot U^{1.41} \cdot u_{rms}^{0.19} \cdot \Lambda_f^{-0.63}, \qquad (22)$$

where the units for $U$ and $u_{rms}$ are m/s, and for $\Lambda_f$ is m. It is important to highlight that this expression was obtained from the experimental dissipation frequency for $\Phi_{uu}|_{\exp}$ only, and not for $\Phi_{ww}|_{\exp}$.

Figures 10 and 11 depict the determined experimental dissipation wavenumber ($k_d|_{\exp} = 2\pi f_d|_{\exp}/U$) and the predicted dissipation wavenumber ($k_d|_{\text{pred}} = 2\pi f_d|_{\text{pred}}/U$), where $f_d|_{\text{pred}}$ was calculated by Eq. 22. As can be inferred from this figure, the predicted dissipation frequencies are considerably similar to the experimental ones.

To model the decay in the dissipation range, a similar equation as the one proposed by Pope [15] (Eq. 17) was

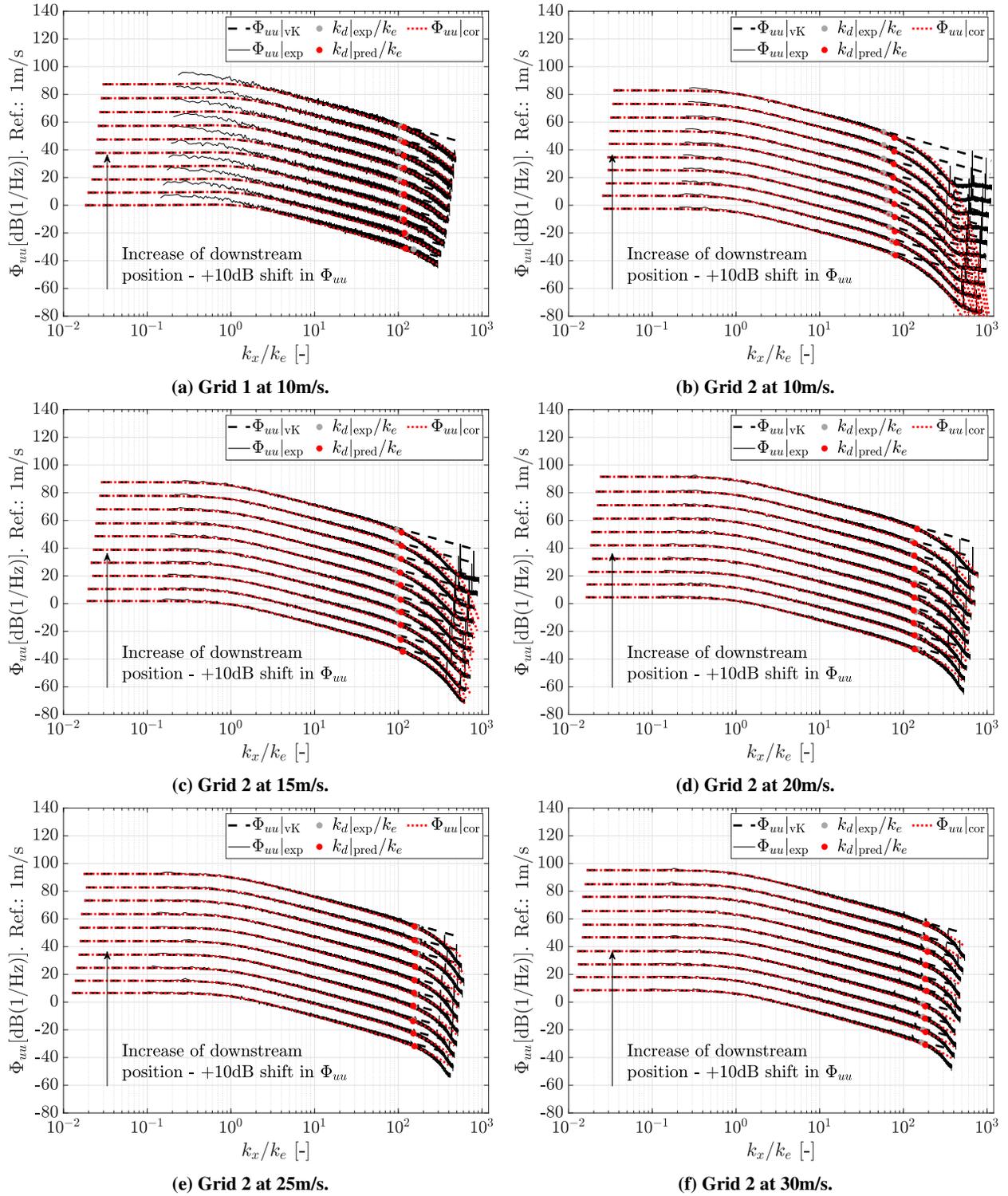

**Fig. 10** Power spectral density of the streamwise velocity for different downstream positions (check Table 2) for grid 1 and 2 for different velocities: dashed line, original von Kármán spectrum $\Phi_{uu}|_{\text{vK}}$ given by Eq 14; continuous line, experimental spectrum $\Phi_{uu}|_{\text{exp}}$; red dotted line, corrected von Kármán spectrum $\Phi_{uu}|_{\text{cor}}$ given by Eq. 24; gray dot, experimental dissipation wavenumber; red dot, predicted dissipation wavenumber given by Eq. 22. The spectrum levels are shifted by 10 dB to facilitate the visualization.

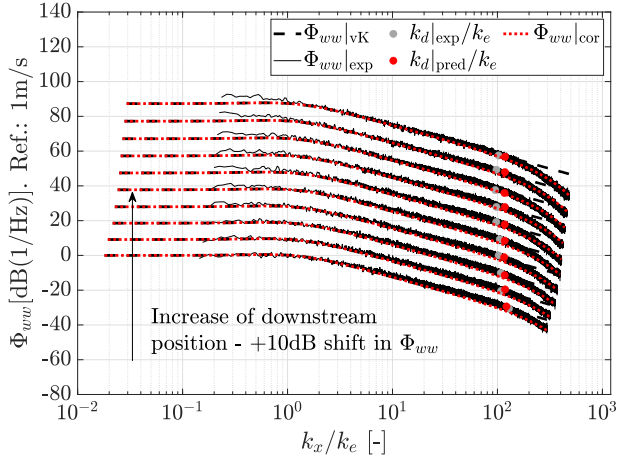
**(a) Grid 1 at 10m/s.**

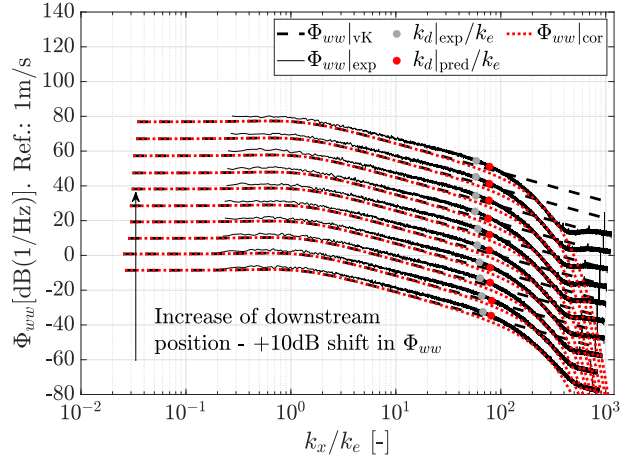
**(b) Grid 2 at 10m/s.**

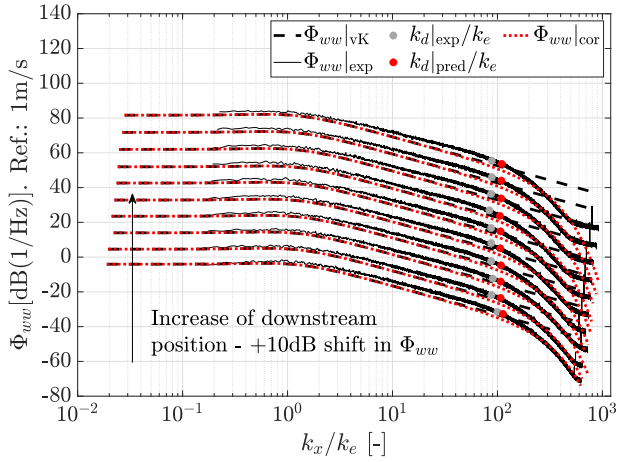
**(c) Grid 2 at 15m/s.**

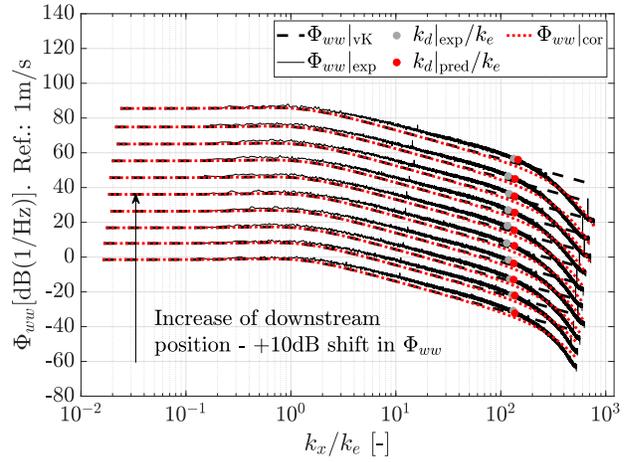
**(d) Grid 2 at 20m/s.**

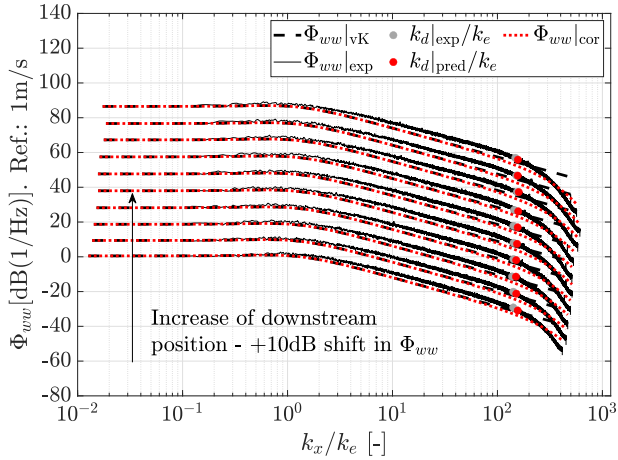
**(e) Grid 2 at 25m/s.**

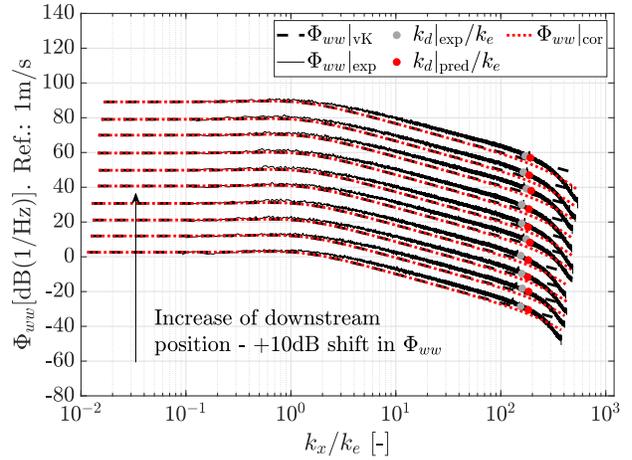
**(f) Grid 2 at 30m/s.**

**Fig. 11 Power spectral density of the streamwise velocity for different downstream positions (check Table 2) for grid 1 and 2 for different velocities: dashed line, original von Kármán spectrum $\Phi_{ww}|_{vK}$ given by Eq 15; continuous line, experimental spectrum $\Phi_{ww}|_{exp}$; red dotted line, corrected von Kármán spectrum $\Phi_{ww}|_{cor}$ given by Eq. 25; gray dot, experimental dissipation wavenumber; red dot, predicted dissipation wavenumber given by Eq. 22. The spectrum levels are shifted by 10 dB to facilitate the visualization.**

used. The dissipation frequency from Eq. 22 was used to compute the Taylor microscale as $\lambda_f = U/f_d$. With $\lambda_f$, the dissipation rate $\epsilon$ (Eq.19) was determined, and consequently the Kolmogorov length scale $\eta$ (Eq. 18). The decay was model as:

$$f_\eta|_{\text{cor}}(k_i, \eta) = e^{-B \cdot (k_i \cdot \eta)^n}, \quad \text{where } B = 3.6 \text{ and } n = 1.5, \tag{23}$$

$$\Phi_{uu}|_{\text{cor}}(k_x) = \Phi_{uu}|_{\text{vK}}(k_x) \cdot f_\eta|_{\text{cor}}(k_x, \eta), \tag{24}$$

where $i$ represents the wavenumber that should be taken into account: if we are predicting the dissipation range for a spectrum as a function of $k_x$, then $k_i = k_x$; if we are predicting the dissipation range for a spectrum as a function of $k_x$ and $k_y$, then $k_i = (k_x^2 + k_y^2)^{0.5}$; and if we are predicting the dissipation range for a spectrum as a function of $k_x$, $k_y$ and $k_z$ then $k_i = (k_x^2 + k_y^2 + k_z^2)^{0.5}$. The constants $B$ and $n$ were determined by fitting the proposed equation to the experimental data from the dissipation frequency up to the frequency correspondent to a spectrum level of -60 dB since for levels lower than -60 dB the hot-wire noise started to play a role.

Figure 10 shows the experimental velocity spectrum $\Phi_{uu}|_{\text{exp}}$, the original von Kármán spectrum $\Phi_{uu}|_{\text{vK}}$ given by Eq 14, and the corrected von Kármán spectrum $\Phi_{uu}|_{\text{cor}}$ given by Eq. 24 with the Kolmogorov length scale obtained from the predicted dissipation frequency (Eq. 22). From this figure, one can see that the dissipation frequency determined with Eq. 22 in combination with the corrected von Kármán spectrum (Eq. 24) predicts the start of the dissipation range quite well. The dissipation decay expression proposed in this study models the dissipation decay observed experimentally with a slight deviation for higher frequencies. Looking closely to the experimental dissipation range, one can notice that an inflection point is observed in the dissipation range decay, changing the initial decay of the curve. Despite this, the proposed correction predicts the dissipation range relatively well.

The same dissipation decay expression applied to the $\Phi_{uu}|_{\text{vK}}$ is applied to $\Phi_{ww}|_{\text{vK}}$, yielding:

$$\Phi_{ww}|_{\text{cor}}(k_x) = \Phi_{ww}|_{\text{vK}}(k_x) \cdot f_\eta|_{\text{cor}}(k_x, \eta). \tag{25}$$

It is important to highlight that the dissipation frequency expression (Eq. 22) as well as the exponential expression for the dissipation range (Eq. 23) were obtained from the experimental data for $\Phi_{uu}|_{\text{exp}}$ only, and not for $\Phi_{ww}|_{\text{exp}}$. In this way, the proposed expressions can be validated using the data for $\Phi_{ww}|_{\text{exp}}$. The dissipation frequency and the decay of the dissipation range should be the same for any energy spectrum considered, as can be inferred from Fig. 1. This is a consequence of the universal behavior presented by the small eddies [15].

Figure 11 shows the experimental velocity spectrum $\Phi_{ww}|_{\text{exp}}$, the original von Kármán spectrum $\Phi_{ww}|_{\text{vK}}$ given by Eq 15, and the corrected von Kármán spectrum $\Phi_{ww}|_{\text{cor}}$ given by Eq. 25 with the Kolmogorov length scale obtained from the predicted dissipation frequency (Eq. 22). From this figure, one can see that the dissipation frequency predicted and the one determined experimentally are similar, even though the predicted dissipation frequency was modeled only with the $\Phi_{uu}$ data. This confirms that the dissipation frequency modeled is quite accurate and it is the same for different energy spectra. The proposed expression for the dissipation range also models relatively well the decay for $\Phi_{ww}|_{\text{exp}}$, presenting the same inflection point as observed for $\Phi_{uu}|_{\text{exp}}$. This confirms the validity of the decay expression for the dissipation range for different energy spectra.

### F. Leading edge noise prediction

The power spectral density of the leading edge noise was predicted using Amiet's formulation presented in Eq. 16, which considers the von Kármán model for $\Phi_{ww}(k_x, k_y)$ as an input. To have a more accurate noise prediction, the dissipation range decay must be applied to the leading edge noise model. According to Pope [15], the energy spectrum $E(k)$ should be multiplied by the dissipation range function $f_\eta$, and then the integral of the resulting expression in relation to $k_z$ has to be taken in order to obtain $\Phi_{ww}(k_x, k_y)$:

$$\Phi_{ww}|_{\text{Pope}}(k_x, k_y) = \int_{-\infty}^{\infty} \frac{E(k) \cdot f_\eta|_{Pope}}{4\pi k^2} \left(1 - \frac{k_z^2}{k^2}\right) \mathrm{d}k_z. \tag{26}$$

The correction proposed in this paper for the von Kármán spectrum considers that the energy spectrum of interest ($\Phi_{uu}$, $\Phi_{ww}$) must be multiplied directly by the dissipation range model (see Eqs. 24 and 25), i.e., the integral of $E(k)$ taking into account the dissipation range expression is not performed. To verify the validity of this procedure, the $\Phi_{ww}(k_x, k_y)$ was obtained in three ways:
- Pope's full formulation: $\Phi_{ww}(k_x, k_y)$ was obtained by performing the integral of $E(k)$ multiplied by Pope's model for the dissipation range, as proposed by Pope [15], see Eq. 26.

- Pope's simplified formulation: $\Phi_{ww}(k_x, k_y)$ was obtained by multiplying $\Phi_{ww}|_{vK}(k_x, k_y)$ by Pope's model for the dissipation range:

$$\Phi_{ww}|_{\text{Pope,simp}}(k_x, k_y) = \Phi_{ww}|_{vK}(k_x, k_y) \cdot f_\eta|_{\text{Pope}}(k, \eta), \qquad (27)$$

where the input $k$ for $f_\eta|_{\text{Pope}}$ was considered as $\sqrt{k_x^2 + k_y^2}$, and the input $\eta$ was computed from the dissipation frequency calculation (Eq. 23).

- Proposed formulation in this study: $\Phi_{ww}(k_x, k_y)$ was obtained by multiplying $\Phi_{ww}|_{vK}(k_x, k_y)$ by the dissipation range model proposed in this paper:

$$\Phi_{ww}|_{\text{cor}}(k_x, k_y) = \Phi_{ww}|_{vK}(k_x, k_y) \cdot f_\eta|_{\text{cor}}(k_i, \eta), \qquad (28)$$

where the input $k_i$ for $f_\eta|_{\text{cor}}$ was considered as $\sqrt{k_x^2 + k_y^2}$, and the input $\eta$ was computed from the dissipation frequency calculation (Eq. 23).

Figure 12 shows a comparison between the original von Kármán spectrum $\Phi_{ww}|_{vK}(k_x, k_y)$ given by Eq. 13, the von Kármán spectrum with Pope's dissipation range considering the full spectrum form $\Phi_{ww}|_{\text{Pope}}(k_x, k_y)$ given by Eq. 26, the simplified form $\Phi_{ww}|_{\text{Pope,simp}}(k_x, k_y)$ given by Eq. 28, and the von Kármán spectrum with the model proposed in this study $\Phi_{ww}|_{\text{cor}}$. The experimental data for grid 2 at 15 m/s at the LE position (Table 4) was used as input for the energy spectrum and the Kolmogorov length scale was determined based on the dissipation frequency model (Eq. 23). As can be seen from this figure, the difference between the full form and the simplified form of the von Kármán spectrum corrected by Pope's dissipation range model is negligible. This demonstrates that multiplying the dissipation decay model by the von Kármán energy spectrum without performing the integration of this model gives the same result. The proposed dissipation model presents a faster decay than the model proposed by Pope, as expected since the proposed model considers the wavenumber to the power of 1.5.

Since $\Phi_{ww}|_{vK}(k_x, k_y)$ can be directly multiplied by the dissipation range expression, the proposed model for the dissipation range was applied to the term $\Phi_{ww}(k_x, k_y)$ of Amiet's leading edge noise prediction model, and the result is shown in Fig. 12. This figure also includes the noise prediction considering the original von Kármán spectrum. The input data for the Amiet's model were the experimental data for grid 2 at 15 m/s at the LE position (Table 4) considering an observer perpendicular to the airfoil at $z = 1.5$ m and an airfoil with a chord and span lengths of 0.2 m and 0.7 m, respectively. When the dissipation range model is taken into account, the leading edge noise decays much more rapidly for high frequencies than predicted without the dissipation model. The difference between the two noise predictions reaches a maximum of 17 dB for the highest wavenumber.

## V. Conclusions

We investigated a nearly isotropic inflow turbulence aiming to determine the flow field characteristics that influence the dissipation frequency in order to model this frequency and the dissipation range for the turbulence energy spectrum. Two passive grids were used and the flow was characterized by hot-wire measurements.

The grids were designed to generate a nearly isotropic turbulence at the airfoil leading edge position, which was confirmed by the measurements. The grids generated a higher streamwise turbulence intensity than predicted, with grid 1 generating around 20.5% and grid 2 around 10.5%. The velocity was mostly uniform at the test section center where the airfoil leading edge is located. The turbulence intensity was mostly uniform in the test section, except at the extremities due to the interaction of the grid turbulence with the open test section shear layer. The lateral length scale was half of the integral length scale, strongly indicating that the flow was nearly isotropic. Furthermore, the velocity energy spectrum matched well with the von Kármán energy spectrum up to the dissipation range since the von Kármán does not take into account the dissipation for the smallest length scales.

We proposed two empirical models in this study: 1. to predict the dissipation frequency; and 2. to predict the spectrum dissipation range based on Pope's model. From the experimental data it was observed that the dissipation frequency depends on the turbulence intensity, the free-stream velocity, and the turbulence length scale. Hence, a relation to compute this frequency as a function of these parameters is proposed. The dissipation range was based on the model proposed by Pope [15] but with a different constant and an exponent was added in order to have a better agreement with the experimental data. The spectrum dissipation range is well predicted by these two models.

The empirical expressions proposed in this paper for the dissipation frequency and the spectrum dissipation range were applied to the Amiet's leading edge noise prediction model. The leading edge noise considering the dissipation range modelling decays much more rapidly for high frequencies than predicted by the von Kármán model. The difference between the leading edge noise prediction using the original von Kármán model and the von Kármán model with the dissipation range correction reaches a maximum of 17 dB for the highest frequency.

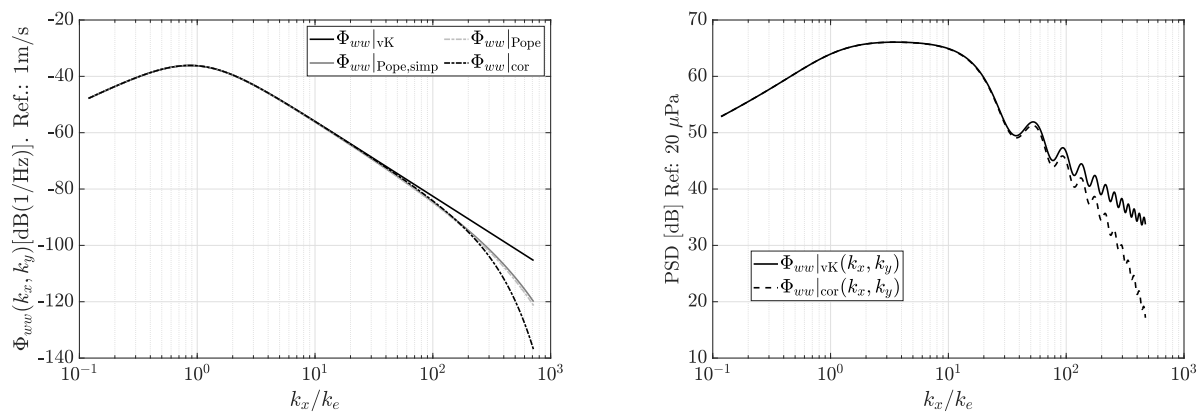

Fig. 12 Left: Comparison between the original von Kármán spectrum $\Phi_{ww}|_{\text{vK}}(k_x, k_y)$, the von Kármán spectrum with the full form of Pope's model $\Phi_{ww}|_{\text{Pope}}(k_x, k_y)$, the von Kármán spectrum with the simplified form of Pope's model $\Phi_{ww}|_{\text{Pope,simp}}(k_x, k_y)$, and the von Kármán spectrum with the model proposed in this study $\Phi_{ww}|_{\text{cor}}$. The experimental data for grid 2 at 15 m/s at the LE position was used as input for the energy spectrum and the Kolmogorov length scale was determined based on the dissipation frequency model. Right: Comparison of the Amiet's leading edge noise prediction model considering the original von Kármán spectrum and the proposed spectrum $\Phi_{ww}|_{\text{cor}}(k_x, k_y)$. The input data for the model were the experimental data for grid 2 at 15 m/s at the LE position (Table 4) considering an observer perpendicular to the airfoil at $z = 1.5$ m and an airfoil with a chord and span lengths of 0.2 m and 0.7 m, respectively.

## Acknowledgments

The authors would like to acknowledge Ing. W. Lette, ir. E. Leusink, S. Wanrooij, and the technicians from the metal workshop of the University of Twente for the technical support. The authors also would like to acknowledge TNO and the Maritime Research Institute Netherlands (MARIN) for the insightful discussions. Part of this research received support from the European Union's Horizon 2020 research and innovation programme under the Marie Sklodowska-Curie grant agreement No 860101.